\documentstyle{ioplppt}                

\input psfig.tex
\begin{document}
\title{Satellite observations of thought experiments close to a black hole}

\author{S. K. Chakrabarti}

\address{{S. N. Bose National Centre For Basic Sciences
        JD Block, Salt Lake, Sector-III, Calcutta-700098, India
        email: chakraba@boson.bose.res.in}}

\begin{abstract}
Since black holes are `black', methods of their identification must necessarily 
be indirect. Due to very special boundary condition on the horizon, the advective flow
behaves in a particular way, which includes formation of centrifugal
pressure dominated boundary layer or CENBOL where much of the infall energy is
released and outflows are generated. The observational aspects  of black holes 
must depend on the steady and time-dependent properties of this boundary layer.
Several observational results are written down in this review which seem to 
support the predictions of thought experiments based on this 
advective accretion/outflow  model. In future, when gravitational waves are
detected, some other predictions of this model could be tested as well.
\end{abstract}
%
%

\noindent Published in Classical and Quantum Gravity, V. 17, No. 12, page 2427,
2000

\maketitle

\section{Introduction}

Black holes are very compact astrophysical objects close to which gravity is 
very strong. Light rays emitted in the vicinity are bent back to itself. Matter
rotating around it or infalling in it, moves with speed close to velocity
of light. The gravity is stronger than $1/r^3$ close to the horizon,
and hence matter with any arbitrary amount of angular momentum would
eventually fall inside it, after some `hesitation' at the centrifugal barrier.
Since black holes are `black' by definition, ways to identify them would 
necessarily be indirect, only the degree of `indirectness' would vary. 
The strong gravity close to the horizon would force matter to act differently
than what is normally seen in weak gravity regions. In this review,
I shall elaborate results of a few recent computations which have the power of 
predictability and then show that observations indicate that the results 
generally agree with these  predictions. Most of these predictions involve accretion processes 
and winds/outflows on which I shall concentrate my review.

Black hole accretion models seem to undergo major changes every twenty years or so.
In 1950s, Bondi [1] accretion model was introduced, which describes adiabatic spherical accretion
on a star. This was readily adopted for black hole accretion 
and was applied to explain quasars and black hole spectra. Since matter
falls in rapidly, its density is very little, and the radiation is generally
that of optically thin matter emitting through bremsstrahlung, not enough to
explain quasar luminosity. It was not until the work of Lynden-Bell [2] the concept of 
accretion {\it disks} which include angular momentum was taken seriously. This
idea matured in works of Shakura \& Sunyaev  [3] who systematically developed the
theory of optically thick, but geometrically thin accretion disks. One of the
major predictions of this model is that black body radiation should be emitted 
with local disk temperature and the resulting multi-colour bump in the spectra
should be a major feature in the accretion disk spectrum. This bump occurs at UV radiation 
when the central black hole is super-massive (with mass around $10^{7-8} M_\odot$)
while it occurs at soft X-rays ($\sim 2-3$keV) when the black hole mass is
around $5-10 M_\odot$. This bump was identified in several AGNs 
[4] and it was immediately understood that at least in some
region of the disk, angular momentum distribution {\it is} Keplerian. However, by 1979
it was already clear that high energy radiation is
emitted as a power-law which must be coming out of Comptonization of 
soft photons by hot electron clouds located nearby [5]. Only through systematic study of 
accretion flows with proper boundary conditions, in the 1990s, we are
sure that this Compton cloud is the inner part of the accretion disk itself 
and is not Keplerian. Extensive properties of these transonic flows are
presented in Chakrabarti [6] which clearly show that centrifugal force
could be strong enough to produce standing shocks when viscosity is low 
(less than a critical value). Complete set of solutions for winds for positive
Bernoulli constant as well as for isothermal flow is also provided.
Meanwhile, Chakrabarti \& Wiita [7], using a single temperature
model, found that the standing shocks in the advective flow may be responsible 
for excess UV radiations in AGN spectra. By 1993 it was already clear that the most general
global solution of accretion flow should contain Keplerian and sub-Keplerian
disks, with or without shock. Quoting Chakrabarti [8]: ``These findings are very significant
as they propose a unifying view of the accretion disks. This incorporates two extreme
disk models into a single framework: for inviscid disks, strong shocks are 
produced, and for disks with high enough viscosity, the stable shock disappears
altogether and angular momentum distribution can become Keplerian." Recent models
of advective accretion flows (including 
ADAF [9] which mimic some of these properties, often for wrong reasons:
some of the shortcomings of ADAF are discussed in Chakrabarti [10], Lu [11] 
and Bisnovatyi-Kogan [12]) are offshoot of this general solution. 

In the next Section, we review these new paradigms of
disks and outflow solutions and show 
that the predictions by these disks are generally verified through recent observations.

\section{Brief Outline of the Advective Accretion Disks}

Since the emergence of the standard Keplerian disk model [3]
in the seventies, purely rotating thick accretion disk models [13]
were built in the eighties, with nearly constant specific angular momentum $\lambda$, 
or where the angular momentum varies with radial parameter in the form of a power-law [14].
These are the predecessors of current advective disk models [6, 15-16]
but the current models self-consistently include 
radial motions as well. Certain old models did attempt to include  radial 
motion in the past [17-18] but global solutions were not found correctly.

An accreting matter must have a significant radial motion, especially close to the
compact object. A simple derivation of the specific binding energy $u_t$ of inflowing matter [19]
$$
u_t=\left[\frac{\Delta}{(1-V^2)(1-\Omega l)(g_{\phi\phi}+lg_{\phi t})}\right ]^{1/2}
\eqno{(1)}
$$
where, $\Delta=r^2-2r+a^2$, $V$ is the radial velocity in the rotating frame, $\Omega$ and $l$ are
the angular velocity and angular momentum, $g_{\mu\nu}$ are the metric coefficients of the Kerr metric.
What is clear is that finiteness of $u_t$ guarantees that $V=1$ when $\Delta=0$, i.e., on the horizon.
Since the sound velocity is less that the velocity of light, the matter must be super-sonic on the
horizon. This is not true for a neutron star, since neutrons stars have hard surfaces and matter
must stop on it. This means that for any generic equation of state, independent of outer 
boundary condition, matter must be supersonic on the horizon and subsonic on a neutron star surface.
Since a supersonic flow must be sub-Keplerian, a black hole accretion must deviate from a standard 
Keplerian disk [14-16]. This may be the case even for a neutron star accretion, especially 
when the inflow passes through a sonic point. Since infall time [$t_{infall} \sim r^{-3/2}$] 
is very short compared with the viscous time to transport angular momentum 
[$t_{visc}=\frac{1}{\alpha} (\frac{h}{r})^{-2}(\frac{r}{v_{Kep}})$] 
the specific angular momentum ($\lambda$) must be almost constant, i.e., the centrifugal force
$\lambda^2/r^3$ grows rapidly compared to the gravitational force $1/r^2$ and
slows down the matter forcing it to be subsonic. Due to general
relativistic effects (where gravity is really stronger than $1/r^2$ very close to the
horizon. For instance, the first correction term comes from $l^2/r^4$.), 
matter recovers from this quasi-stagnant condition, passes through
another sonic point and finally enters into the black hole supersonically.
It is easy to show that the general relativity introduces one extra sonic point
(inner sonic point) close to the horizon. This is discussed in detail in 
Sec. 5.1 of [15]. Chakrabarti ([14-16, 19]), solves and classifies all
possible ways inviscid and viscous matter can fall inside a black hole. No new solution
has been found since then.

The temporary stagnation of matter due to centrifugal force (at the so-called centrifugal
force driven boundary layer of a black hole, or CENBOL), especially true when viscosity
is low [14], at a few to a few tens of Schwarzschild radii is of
extreme importance in black hole physics. Slowed down matter would be hotter
and would emit harder radiations. If viscosity is high, the Keplerian disk extends (from outside)
to distances very close to the black hole, perhaps as close as the marginally stable orbit
($3R_g$, $R_g$ being the Schwarzschild radius.). If viscosity is low, the inner edge may
recede outwards to a few tens to a few hundreds of Schwarzschild radii. This is because
angular momentum must be transported outward by viscosity and smaller viscosity
transports slowly taking longer path length to reach a Keplerian disk. Another major point
which must be emphasized is that in order that matter falls to black hole steadily and at the
same time forming a disk, the specific angular momentum at the inner sonic point 
should be between marginally stable and marginally bound values (i.e., between $3.67GM/c$ and $4GM/c$).
Thus, if matter comes with a large angular momentum, viscosity must work its way to ensure
that matter losses most of it before entering a black hole. When the viscosity is so high that 
the specific angular momentum rapidly decreases with decreasing distance,
the centrifugal barrier weakens and the stagnation region disappears [15-16]. In this 
case, the sub-Keplerian region is confined roughly between $3R_g$ and $R_g$ only.

When the stagnation region forms {\it abruptly} in a fast moving sub-Keplerian flow,
a shock is said to have formed. Depending on the parameter space [14-16]
this shock may or may not be standing at a given radius. It may form and propagate to
infinity [20], or may just oscillate [21-23] or may stand still [22, 24]. In any case, generally shock 
forms and it decidedly affects the nature of the spectra
manifesting itself through propagating noise, or quasi-periodic oscillation, or
steady state spectrum, or the formation of the quiescence state. There are 
overwhelming evidences that
these shocks form in the advective accretion flows, details of 
which would be presented below. ADAF models are highly viscous and shocks 
are absent [9].

For a neutron star, accretion flow may be entirely sub-sonic and deviation from a 
Keplerian disk is not essential except in the narrow boundary layer where 
the rotational motion of the accreting matter must adjust to the rotational 
motion of the star and where the radial velocity must go down to zero. This is 
true when magnetic field is absent or very weak. In presence of a 
strong magnetic field, matter is stopped way ahead of the
stellar surface and is bent back along the field line. If the flow deviates from
a Keplerian disk, the ram pressure $\rho v_r^2$ is higher and some matter can penetrate
radially through the field line and directly hit the surface when the field is
weaker. If this flow does become supersonic and sub-Keplerian at any point, 
it must become subsonic just before hitting the surface [24].

In the centrifugal barrier dominated boundary layer of black holes
and neutron stars, winds/outflows may form [14]. This has been quantified very recently
[10, 25] where the ratio of outflow to inflow rates has been computed as
functions of the compression ratio of matter at the CENBOL. (Detailed analysis 
with and without shocks are in Das \& Chakrabarti [26]). Chakrabarti [15] pointed out
that in presence of magnetic field, the outflow or jets coming out of the post-shock
region could be blobby as seen in the black hole candidate GRS1915+105. Subsequently, 
it was pointed out (see below, [33]) that blobby flows may form even without 
a magnetic field. This phenomenon of outflow is more significant for a black 
hole than for a neutron star. Neutron stars are known to have magnetic
fields and therefore most of the matter is stopped by it (unless the field is so 
weak that sub-Keplerian matter with larger ram-pressure penetrates it easily, 
directly hitting the surface). Matter moves along the
field lines to the polar region and most of the winds come from that region only.
On the contrary, black holes may not anchor magnetic fields. Most of the
outflows can thus form at the CENBOL itself, due to its high outward pressure
gradient force, and due to the high temperature. However, it is unlikely that
purely hydrodynamic acceleration could cause a `super-luminal' jet to form [47].
Since in the low-luminosity hard state CENBOL is hotter, the ratio 
$R_{\dot m} = \frac{{\dot M}_{out}}{{\dot M}_{in}}$ should be higher in hard states
than in soft-states (when the CENBOL is cooler). 

There is another region of the outflow: at the boundary between the
Keplerian and the sub-Keplerian matter. As Chakrabarti et al. [27, 28] showed
through extensive numerical simulation, angular momentum distribution
at this transition region is often super-Keplerian and therefore the outward centrifugal
force is very strong. Thus, a good deal of centrifugally driven outflow
at this region cannot be ruled out. In case of neutron stars, because of the
same reason, outflows are possible not only in polar caps, but at the transition
radius as well. If the magnetic field is non-aligned with the rotational axis,
the Coriolis force on both sides of the disk would be slightly different and the
QPO frequencies would split. Such behaviours have been observed in several
neutron stars [29] and these essentially verify outflow solutions from transition regions [27-28].

\section{Schematic pictures of accretion and outflow processes}

As mentioned in the Introduction, with the advent of more detailed and accurate
observations which necessarily demand more accurate solutions for proper explanations,
it has become necessary to revise models of accretion and outflows. Recently, it is understood
that black holes have two major types of spectra [30-32]: one is the soft state 
where most of the power is emitted in soft radiation ($\sim 2-10$keV), 
and the other is the hard state, where most of the power is 
emitted in the hard radiation ($\sim 10-50$keV). Up until 5-6 years ago, it was believed that
the hard radiation is the result of Comptonization of soft photons by `Compton Clouds' 
floating around the disk or by hot corona. Chakrabarti and Titarchuk [30], for the first time
pointed out that the so-called `Compton Cloud' is the blotted-out inner edge of the 
Keplerian disk itself! This has brought black hole accretions to a new
paradigm. Today, this picture is universally adopted in most of the models of accretion flows. 
Figure 1a shows the schematic picture of accretion and outflows in the so-called `hard' state 
and Fig. 1b shows the schematic picture in the so-called `soft' state. The difference is that
in the hard state, Keplerian flow rate could be very small while the sub-Keplerian
rate could be very high [30]. In the soft state, it is the opposite. The de-segregation 
of matter into these two types of rates are believed to be due to the fact that 
the flow closer to the equatorial plane is likely to be more viscous and with larger
Shakura-Sunyaev parameter $\alpha$, and therefore is likely to be Keplerian. However,
flows away from the equatorial plane may have smaller $\alpha$ and therefore they deviate
from a Keplerian disk farther out. Contribution to this sub-Keplerian may also have
come from wind accretion. This sub-Keplerian flow may form shocks at around 
$10-20R_g$. In the soft state, shocks may be nominally present, but would be
cooler due to Comptonization, and would be as good as non-existent. The combined picture
in these two states was shown in Chakrabarti \& Titarchuk [30]. As far as the
wind and outflow formation goes, the result strongly depends on the  compression ratio ($R$) 
of the flow at the shock as discussed in Chakrabarti [10, 25] where an analytical expression 
for the ratio $R_{\dot m}$ of the outflow (assumed isothermal) to inflow rates is provided. 
When the compression ratio is unity, i.e., in the shock-free case (i.e., soft state),
the outflow rate is negligible. Thus, in soft states, no outflow is expected.
In the hard state, shock is strong, and the outflow may be significant. In the
intermediate state outflow rate is highest (see below). In Das \& Chakrabarti [26] 
it was shown that outflow rate should generally depend on inflow rates as well  
and the ratio is very high for low accretion rates.  

\begin{figure}
\vbox{
\vskip 0.0cm
\hskip -0.0cm
\centerline{
\psfig{figure=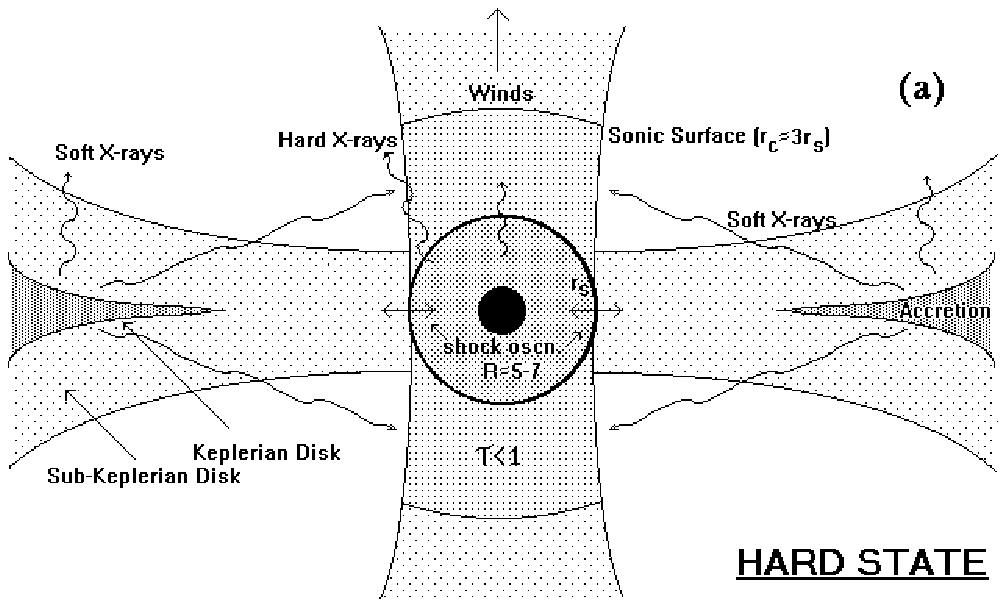,height=5truecm,width=7truecm}}}
\begin{verse}
\vspace{0.0cm}
\end{verse}
\end{figure}

\begin{figure}
\vbox{
\vskip -1.5cm
\hskip -0.0cm
\centerline{
\psfig{figure=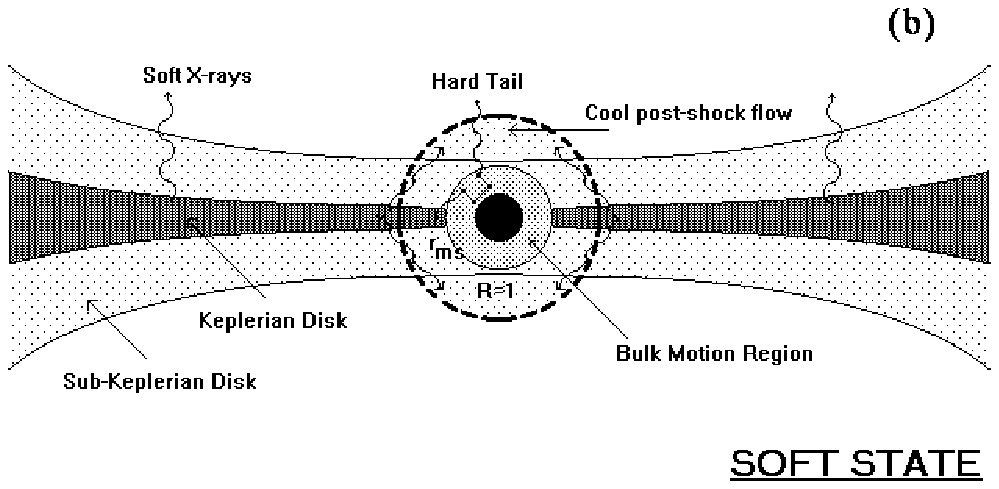,height=5truecm,width=7truecm}}}
\begin{verse}
\vspace{0.0cm}

\end{verse}
\end{figure}

\begin{figure}
\vbox{
\vskip -1.5cm
\hskip -0.0cm
\centerline{
\psfig{figure=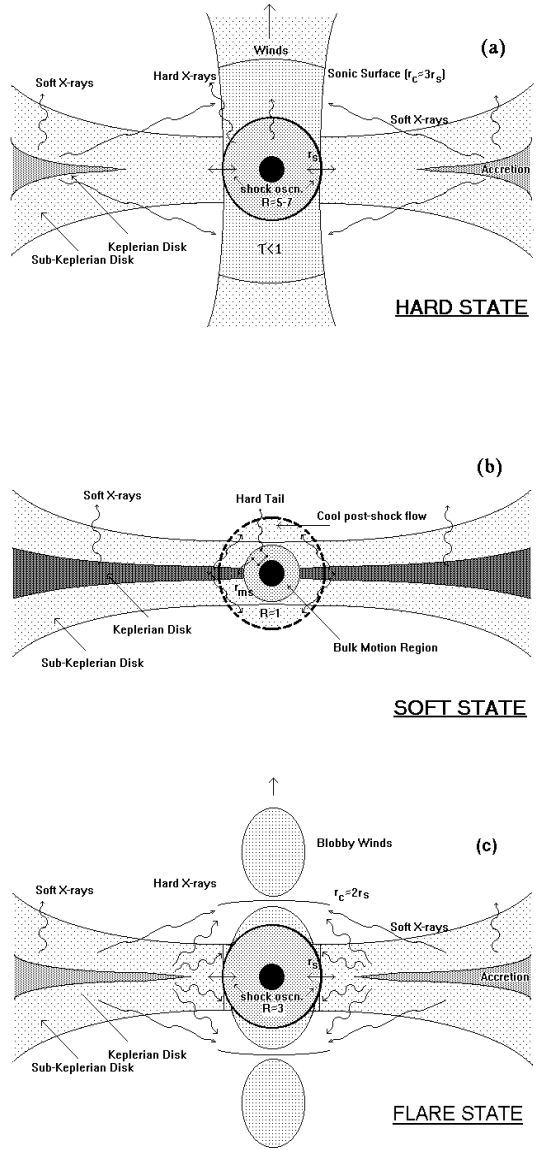,height=5truecm,width=7truecm}}}
\begin{verse}
\vspace{0.0cm}
\caption{ Schematic picture of accretion and outflow around a black 
hole in hard (upper), soft (middle) and intermediate or flaring 
(lower) states. These states are distinguished by compression ratio 
of the shocks which in turn is determined by the accretion rates and 
radiation transfer properties of the flow. Outflow rate is determined 
by the compression ratio. Hence formation of the outflow is strongly 
linked to the states of the black hole spectrum [25, 33].}
\end{verse}
\end{figure}

For intermediate compression  ratio $R$, outflow rate is the highest, and if other conditions
(namely, conditions for shock oscillations) are satisfied, X-rays may exhibit quasi-periodic oscillations
[25] and rapid switch from harder and softer radiations are exhibited in so-called 
burst phases. This is schematically shown in Fig. 1c. In this case, the outflow would be blobby 
[33-34]. More detailed discussions would be made in the next Section.

A curious observation of generation of blobby jets was made, purely from theoretical point of view,
in the context of anti-correlation found in the X-ray and radio fluxes in the black hole candidate
GRS1915+105 (see, \S 4.7-4.8 of [15]). When magnetic field is dragged in by the accretion disks,
it could be sheared and amplified in the post-shock region till an equipartition is reached. A strong
field in a hot gas feels magnetic tension (`rubber-band effect') and is contracted catastrophically
evacuating the post-shock flow, i.e., the inner part of the accretion disk. Quoting [15]:
`` Formation of coronae in an accretion disk is not {\it automatic}, i.e., presence of magnetic field
inside a disk does not automatically imply that a hot magnetic coronae would form. The formation requires
the ability of the accretion disks to anchor flux tubes inside the disk. This means that the disk
should have an internal structure akin to the solar interior and the entropy must increase
outwards. If the entropy condition is proper, the coronae would form, otherwise it would come out of the
disk as a whole without causing any random flare. In former case, there would be sporadic flaring events on the disk
surface as in the case of the sun, whereas in the latter case, the collapse of fields 
in the funnel would cause destruction of the 
inner part of the disk and formation of blobby radio jets. Detailed observation of GRS1915+105 shows these
features [35]. Since the inner part of the accretion disk could literally disappear by this magnetic 
process, radio flares should accompany reduction of X-ray flux in this objects. Since the physical 
process is generic, such processes could also be responsible for the formation of jets in active
galaxies and similar anti-correlation may be  expected, though time delay effects are to be incorporated
for a detailed modeling.'' In the absence of the magnetic field, periodic evacuation by 
oscillating shocks can also take place if the flow parameters are `wrong' (see, [22] 
for details). This phenomenon may have been observed very convincingly in GRS1915+105.

\section{Predictions of Advective Accretion and Winds}

Several interesting ways advective accretion can manifest themselves. 

\subsection{Manifestation of Sub-Keplerian motion on a large scale}
Sub-Keplerian flows rotate slower, and velocity predicted from Doppler shifted disk emission lines 
would correspond to a higher Central mass. Chakrabarti [36] pointed out that the disk around M87 
contained shocks and therefore the flow must be sub-Keplerian. As a result, the mass of the central object is
around $4\times 10^{9} M_\odot$ rather than $2 \times 10^9M_\odot$ as predicted by Harms et al. 
and Ford et al. (see, [36] for references) purely based on Keplerian motion. 
The fact that shock-ionization causes the emission processes on M87 disk has been 
stressed by several others [37]. It is expected that most of the derived masses of 
the black holes at galactic centres are seriously underestimated because of the presence 
of the sub-Keplerian and non-dissipative flows [15].

\subsection{Deviation from a Keplerian Disk}
Traditionally, it is assumed that the inner edge of a Keplerian disk extends till $3R_g$. However,
after the development of the advective disk models [14-16, 30]
this traditional picture has changed. It was predicted that the inner edge would be closer
to $\sim 10-20 R_g$ where the CENBOL should form. There are overwhelming evidence today that
this is indeed the case [38-40]. This sharply contrasts predictions of ADAF model which often 
models the inner edge a distance of several tens of thousands (even at hundreds of 
thousands) of Schwarzschild radii! 

\subsection{Power-law hard radiation in very high states}
Chakrabarti \& Titarchuk [30] pointed out that when the accretion rate is relatively high,
the electrons in the sub-Keplerian region become cooler and this region becomes practically 
indistinguishable from that of Keplerian disk. However, very close to the black hole horizon,
matter moves with almost velocity of light and deposits its bulk momentum onto the photons
thereby energizing these photons to very high energy forming a power-law. This power-law
has now been the hall-mark of all the known black holes [42]. The success of this model 
crucially hangs on the transonic flow solution  which utilizes the fact that the inner 
boundary condition is independent of the history of incoming matter.

\subsection{Explanation of quasi-periodic oscillations from black hole candidates}
X-rays from galactic black hole candidates often show persistent oscillations
which are quasi-periodic in nature. While the standard accretion disks do not show
any signatures of oscillations, the advective disk solutions do, especially in the
X-ray emitting regions. There are two general causes of these oscillations. One is 
due to the inability of the steady shocks to form in the sub-Keplerian region when the
Rankine-Hugoniot relation is not satisfied [22]. In this case,
the disk is periodically evacuated and strong wind and outflows are formed. The other reason 
for oscillation is the resonance between the cooling time and the infall time. When the
black hole accretion rate changes from hard to soft states, the accretion rate, and therefore the
cooling times change. When this roughly agrees with the infall time from the post-shock
region, the otherwise steady shock starts oscillating with a period roughly comparable to the
infall time [21]. The observation of these oscillations during the transition of states 
is the triumph of the advective disk model [43] ADAF models, unlike our solutions, 
do not allow shock formation [9] and therefore cannot explain the quasi-periodic 
oscillations at all.

\subsection{Formation of outflows and modification of spectral properties}
Just as Parker's solution of Solar wind is the outgoing solution of 
accretion flows, there are outgoing transonic solutions which are counterparts of 
transonic accretion flows onto black holes [44]. These outflows may 
or may not have shocks, but they invariably have positive binding energy at a large distance.
ADAF solution recently came to the same conclusion (that Bernoulli
parameter should be positive [45]). However, when the outflow is accelerated by external agency
such a conclusion is wrong as has been recently shown by Chattopadhyay and Chakrabarti 
[46]. They showed that even matter with negative initial binding energy could 
be pushed to infinity (large distance) when radiation momentum is systematically 
deposited on the flow! This again shows that ADAF models are grossly incomplete. 
Recently, to overcome some of the shortcomings 
ADAF has also tried to produce self-similar outflows (ADIOS [47])
Being self-similar, jets are to come out all over the disk,
as opposed to regions close to the black holes as predicted by our advective disk
solutions. Recent high resolution observations of jets in M87 strongly suggest that 
they are produced within a few tens of Schwarzschild radii of the horizon [48], 
strongly rejecting ADAF and ADIOS models for the outflows,
which has no special length scale at these distances. 

Given an inflow rate, one is now capable of computing the outflow rate
when the compression ratio at the shock surface is provided [25, 49]
This solution naturally predicts that the outflow must form at the CENBOL. The
Globally complete Inflow-Outflow Solutions (GIOS) were also found [49]. 
In presence of winds, the spectra is modified: outflows take away
matter and reduces the number of electrons from the post-shock (CENBOL) region
and thus they are cooled more easily by the same number of soft photons emitted 
by the pre-shock Keplerian flow. This makes the spectra much softer [50]
in presence of winds. 

\subsection{Quiescence states of black holes}
Chakrabarti [10] and Das \& Chakrabarti [26], pointed out that in some regions of the
parameter space, the outflow could be so high that it evacuates the disk and forms
what is known as the quiescence states of black holes. A well known example is the 
starving black hole at Sgr A* at our galactic centre whose mass is
$\sim 2.6\times 10^6 M_\odot$ and the accretion rate is supposed to 
be around $\sim 10^{-5} M_\odot$ yr$^{-1}$ which is much smaller compared to the 
Eddington rate. Quiescence states are also seen in stellar mass black hole 
candidates such as A0620-00 and V404 Cygni. Another way of producing these states 
is to  use well known viscous instability in an accretion disk as used  models of dwarf-novae
outbursts. This was mentioned in Chakrabarti \& Titarchuk [30].

\subsection{On and Off-states during QPOs in black holes}
The black hole candidates GRS1915+105 displays a variety of behaviour: usual high frequency QPOs are
frequently interrupted by low frequency oscillations. It has been recently suggested [25]
that these low frequency oscillations are due to periodic cooling of the 
subsonic region of the outflow (also known as the extended corona). While the QPO frequency can be
explained by the shock oscillations, the periodic cooling is explained by the duration in which
extended corona becomes optically thick. It is possible that outflows are slowed down by this 
process and the matter falls back to CENBOL, extending the duration of the `on' state 
(which, if exists, is found to be comparable to the duration of the `off' state [52-53]). 
The QPO frequency does evolve during this time scale and it is suggested that this is due to
the steady movement of the inner edge of the Keplerian accretion disk [54]
or the steady movement of the shock itself in viscous time scale [34].
Generally speaking, shocks may show large scale oscillations [21-22] but the wind plays a 
vital role in intercepting larger number of soft photons, characteristics of soft state. 
The correlation between the duration of the off-state and the QPO frequency has been
found to agree with the observations [34].

\subsection{Relationship of outflows and the black hole states}
It has become apparent that outflow rates are the highest when the compression ratio $R$ of the gas 
is intermediate ($R\sim 3$; [10,25,33,49]. When the compression ratio is unity 
(i.e., no compression other than the simple geometrical ones) no outflow is expected from these models. 
Similarly, when the compression ratio of the gas is very high, the outflow rate relative to inflow goes down. 
Figure 2 shows the variation of the outflow rate with the compression ratio (taken from [33]). In the 
soft state, when the post-shock region is cool [30], shock is as good as non-existent. 
In this case the outflow rate is also very low since $R\sim 1$. The outflow
rate should be very little. On the other hand, when viscosity and accretion rate are so low that 
very strong shock forms ($R\sim 5-7$; [14, 20]), then it would be a hard state
without a QPO. When the accretion rate is higher such that some sort of resonance condition is satisfied [21]
then the shock would oscillate and form QPOs. Higher
accretion rate softens the shock by cooling the post-shock flow and it becomes of intermediate strength [30].
Thus, the speculation is that the QPOs which are due to shock oscillations
should not be formed in soft states, and should be in hard states only if the resonance condition is
satisfied. Similarly, outflows should not be formed in soft state, and high outflows (relative to 
inflow) should form only when the shock compression ratio is intermediate.  Recently, from observational
point of view this picture has gained support [55].
\begin{figure}
\vbox{
\vskip -1.5cm
\hskip -0.0cm
\centerline{
\psfig{figure=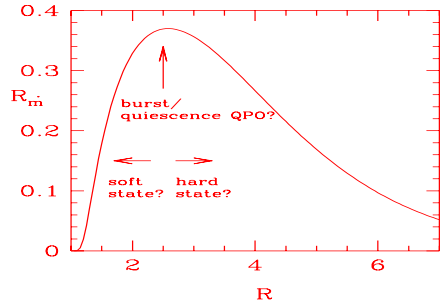,height=8truecm,width=10truecm}}}
\begin{verse}
\vspace{0.0cm}
\caption{Variation of the ratio of the outflow rate to the inflow rate $R_{\dot m}$ with the compression ratio $R$ of the 
shock. It is believed that in soft states, $R\sim 1$ and thus the outflow rate is also very low. In hard states, shocks
are stronger and compression ratio is higher. Here, $R_{\dot m}$ is high but not maximum. For intermediate $R\sim 3$, 
$R_{\dot m}$ is highest, outer sonic point of the flow $r_c$ is closest to the black hole. In this case, the outflow may be
periodically cooled and it may therefore be blobby [33]. }
\end{verse}
\end{figure}

\subsection{Effect of advective disks on gravitational wave}
It is well known that compact binary companions loss angular momentum due to emission of
gravitational waves. Study of gravitational wave emission from binary systems has 
received a significant boost in recent years because of the realization 
that the detection  of gravitational waves would directly identify such coalescing bodies.
Laser Interferometric Gravitational Wave Observatory (LIGO) and
Laser Interferometer Space Antenna (LISA) project instruments are
being constructed  to achieve these goals (see, Thorne [56] for a review).

In binary systems composed of only neutron stars and stellar black holes,
these computations are adequate. However, when studying effects around
a massive black hole which is assumed to be present in centers
of many galaxies, one needs to consider an additional effect -- the effect of an accretion disk.
This was first pointed out by Chakrabarti [57-58].
It is widely believed that galactic centers are endowed with massive black holes
and in order to explain the observed luminosity from a galactic core, one needs to supply
matters ranging from  a few hundredth or solar masses to a few solar masses per year, some of
which may be in the form of stars.  Some of the stars could be
compact, namely, neutron stars and stellar mass black holes which
orbit the massive ones at the same time gradually spiraling in 
towards the center due to loss of angular momentum by gravitational waves.

Chakrabarti [57] pointed out that the accretion disks close to the
black hole need not be Keplerian and it would affect the
gravitational wave properties. The radiation pressure
dominated disks can be super-Keplerian in some regions
(the very basis on which thick accretion disk models are made) which would
transfer angular momentum to the orbiting companion and in some
extreme situations, can even stabilize its orbit from coalescing
any further. This was later verified by time-dependent numerical simulations [59]
When one considers the more general solutions of viscous, transonic, accretion disks [14-16]
one finds that the angular momentum distribution close to the black hole
could be sub-Keplerian as well, depending upon the viscosity
and the angular momentum at the inner boundary of the disk.
Assume that a companion of mass $M_2$ is in an instantaneous
circular Keplerian orbit around a central black hole of mass $M_1$.
This assumption is justified, specially when the orbital radius is
larger than a few Schwarzschild radius where the energy loss per orbit
is very negligible compared to the binding energy of the orbit.
The rate of loss of energy $dE/dt$ in this binary system with orbital
period $P$ (in hours) is given by [60-61];
$$
\frac{dE}{dt}=3 \times 10^{33} (\frac {\mu}{M_\odot})^2
(\frac{M_{tot}}{M_\odot})^{4/3} (\frac{P}{1 hr})^{-{10}/{3}} {\rm ergs\ sec^{-1}},
\eqno{(2)}
$$
where,
$$
\mu=\frac {M_1 M_2}{M_1+M_2}
$$
and
$$
M_{tot}=M_1+M_2.
$$
The orbital angular momentum loss rate would be,
$$
R_{gw}=\frac{dL}{dt}|_{gw}=\frac{1}{\Omega} \frac{dE}{dt}
\eqno{(3)}
$$
where $\Omega=\sqrt{G M_1/r^3}$ is the Keplerian angular velocity of the
secondary black hole with mean orbiting radius $r$. The subscript `gw'
signifies that the rate is due to gravitational wave emission.
In presence of an accretion disk co-planer with the
orbiting companion, matter from the disk (with local
specific angular momentum $l(r)$) will be accreted onto the companion
at a rate close to its Bondi accretion rate [1, 63]:
$$
{\dot M}_2=\frac{4\pi {\bar \lambda} \rho (GM_2)^2}{(v_{rel}^2+a^2)^{3/2}}
\eqno{(4)}
$$
where, $\rho$ is the density of disk matter, ${\bar \lambda}$ is
a constant of order unity (which we choose to be $1/2$ for the rest
of the paper), $v_{rel}=v_{disk}-v_{Kep}$ is the relative
velocity of  matter between the disk and the orbiting companion.
The rate at which the angular momentum of the companion is changed
due to Bondi accretion will be [57],
$$
R_{disk}=\frac{dL}{dt}|_{disk}={\dot M_2} (l_{Kep} (x) -l_{disk} (x) )
\eqno{(5)}
$$
Here, $l_{Kep}$ and $l_{disk}$ are the local Keplerian and disk
angular momenta respectively. The subscript in the left hand
side signifies the effect is due to the disk. If some region of the 
disk is sub-Keplerian ($l_{disk}<l_{Kep}$), the effect 
of the disk would be to reduce the angular momentum of the
companion further and hasten coalescence. This has now been demonstrated with
actual solutions [58] and the nature of the variations of the
signals are presented in Chakrabarti [63].

In order to appreciate the effect due to intervention of the
disk, we consider a special case where, $M_2 <<M_1$ and $l_{disk}<<l_{Kep}$.
In this case, $\mu \sim M_2$ and $M_{tot}\sim M_1$. The ratio $R$ of these two rates is,
$$
R=\frac{R_{disk}}{R_{gw}}=1.518\times 10^{-7} \frac{\rho_{10}}{{T_{10}}^{3/2}}
{x^4}{M_8}^2
\eqno{(6)}
$$
Here, $x$ is the companion orbit radius in units of the Schwarzschild radius of the primary,
$M_8$ is in units of $10^8 M_\odot$, $\rho_{10}$ is the density in units
of $10^{-10}$ gm cm$^{-3}$ and $T_{10}$ is the temperature of the
disk in units of $10^{10}$K. It is clear that, for instance, at $x=10$,
and $M_8=10$, the ratio $R\sim 0.015$ suggesting the effect of the
disk could be a significant correction term to the general relativistic
loss of angular momentum. In the above example, both the disk and the
gravitational wave work in the same direction in reducing angular
momentum of the secondary. Alternatively,  when $l_{disk} > l_{Kep}$
they act in opposite direction and may slow down the loss of angular
momentum [57, 59]. In either case, the ratio $R$ is independent
of the mass of the companion black hole, as long as $M_2 <<M_1$.

Figure 3 shows the effect of the presence of an accretion disk
on the gravity wave pattern in a binary black hole system
consisting of two black holes with mass $10^8 M_\odot$ and $10^6 M_\odot$.
The solution for the disk quantities is obtained using hydrodynamic equations
in presence of heating and cooling [16] for $\gamma=5/3$, $\alpha= 0.02$, $f=0.0$,
$x_{in}=2.3$, and $l_{in}=1.7$ ($\alpha$ is the Shakura-Sunyaev
viscosity parameter, $f$ is a constant cooling efficiency factor [33], $x_{in}$ is the 
inner sonic point of the flow, $l_{in}$ is the angular momentum at the inner sonic point.).
Accretion rate ${\dot M}= 1000 {\dot M}_{Edd}$ is chosen to enhance the effect. When ${\dot M}$
is reduced, as is appropriate for any advective disk, the effect
is proportionately smaller. The `chirp'
profiles as functions of real time are compared (results at only the last few
Schwarzschild radii are shown).
When the sub-Keplerian disk is included, the companion falls
more rapidly due to enhanced loss of angular momentum.

\begin{figure}
\vbox{
\vskip 0.0cm
\hskip -0.0cm
\centerline{
\psfig{figure=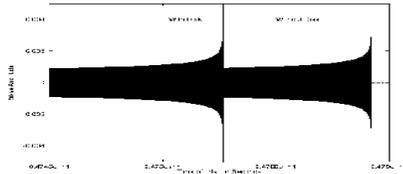,height=8truecm,width=10truecm}}}
\begin{verse}
\vspace{0.0cm}
\caption{
Effect of the presence of an advective accretion disk
on the gravity wave pattern in a binary black hole system
consisting of two black holes with mass $10^8M_\odot$ and $10^6 M_\odot$.
See text for the parameters used. The `chirp' profiles as functions of
real time are compared (profiles in the last few Schwarzschild radii are shown).
When the sub-Keplerian disk is included, the companion falls
more rapidly due to the loss of angular momentum enhanced by combined effects. }
\end{verse}
\end{figure}

One of the exciting predictions of this scenario is that since the spectrum of an
accretion disk contains a large number of informations (e.g. mass of the central black holes,
distance of the black hole, accretion rate, and viscosity parameter) a simultaneous 
observation of the electromagnetic spectrum from the disk and the gravity wave spectrum
(which also must depend on those parameters, except possibly the distance) 
should tighten the parameters very strongly. This issue is being investigated and will be 
reported shortly.

Recently, ADAF model of  extremely low accretion rate was used to repeat 
these computations, and not-surprisingly, no significant change 
in gravitational signal was found [9]. If little matter is accreting,
it is as good as having no accretion disk at all. ADAF model ideally
valid for zero accretion rate systems and thus not found to be true in any of 
the  realistic accreting systems. As discussed earlier,
evidences of the oscillating shocks, bulk motions etc. (which are absent in ADAF) 
are abundant. Thus it is likely that the gravitational signal 
{\it would be} affected exactly in a way computed by [58].

\section{Concluding remarks}

Ten years ago, when globally complete advective disk solutions were 
first introduced which included shocks around black holes, the community
did not take these solutions seriously. Today, with the advent of accurate observational
techniques, including those with high time resolutions,
there is hardly a single paper in the literature which 
does not use the advective disk models. We have already presented 
several such observations and found that everyday there are more
evidences for these solutions. 

So far, most of these works have been carried out using Schwarzschild
black holes. However, black holes should, in general, have some angular 
momentum. Not enough evidence has been obtained, or to be precise, sought,
of rotation of a black hole. This deficiency is partly due to the 
fact that the effect of rotation is manifested only extremely close to the black hole
and in most of the cases, the signatures expected from a rotating (Kerr) black hole,
could be modeled often equally satisfactorily, by a Schwarzschild black hole  
with a different set of accretion disk and black hole parameter. 
In future, the task would be to separate astro physically significant 
effects which depend on Kerr parameters only.

\ack
The author thanks DST for a partial support for the project `Analytical and numerical
studies of astrophysical flows around black holes and neutron stars' and the organizers
to partly supporting his trip to Samarkhand and covering the local hospitalities.

\end{document}